\documentclass{article}

\usepackage{arxiv}

\usepackage[utf8]{inputenc} 

\usepackage{amsfonts}       
\usepackage{nicefrac}       
\usepackage{microtype}      
\usepackage{lipsum}

\usepackage{color}
\usepackage{listings}
\usepackage{booktabs}
\usepackage{array}
\usepackage{graphicx}
\usepackage{longtable}
\usepackage{subfigure}

\usepackage{cite}
\usepackage{url}
\usepackage{fancyhdr}

\usepackage{soul}

\usepackage{float}
\usepackage{listings}
\usepackage{mdwmath}
\usepackage{mdwtab}
\usepackage{multirow}
\usepackage{multicol}
\usepackage{rotating}
\usepackage{setspace}
\usepackage[utf8]{inputenc}
\usepackage{lineno}
\usepackage{listings}

\usepackage{mdwmath}
\usepackage{mdwtab}
\usepackage{multirow}
\usepackage{multicol}
\usepackage{array}
\usepackage{booktabs}

\usepackage{enumitem}
\usepackage{xspace}
\usepackage[export]{adjustbox}
\usepackage{graphicx}
\usepackage{color,soul}
\usepackage{rotating}
\usepackage{setspace}
\usepackage{amsmath} 
\usepackage{amssymb}
\usepackage{float}
\usepackage{xcolor}

\usepackage{hyperref}
\usepackage{url}

\title{AgentArcEval: An Architecture Evaluation Method for Foundation Model based Agents}

\author{Qinghua Lu\textsuperscript{1}, Dehai Zhao\textsuperscript{1}, Yue Liu\textsuperscript{1}, Hao Zhang\textsuperscript{1}, Liming Zhu\textsuperscript{1},\\ \textbf{Xiwei Xu\textsuperscript{1}, Angela Shi\textsuperscript{2}, Tristan Tan\textsuperscript{2}, Rick Kazman\textsuperscript{3}}\\
\textsuperscript{1} Data61, CSIRO, Australia \\
\textsuperscript{2} Empathetic AI, Australia \\
\textsuperscript{3} University of Hawaii, Honolulu, HI, USA
}

\begin{document}

\maketitle

\begin{abstract}
The emergence of foundation models (FMs) has enabled the development of highly capable and autonomous agents, unlocking new application opportunities across a wide range of domains. Evaluating the architecture of agents is particularly important as the architectural decisions significantly impact the quality attributes of agents given their unique characteristics, including compound architecture, autonomous and non-deterministic behaviour, and continuous evolution. However, these traditional methods fall short in addressing the evaluation needs of agent architecture due to the unique characteristics of these agents. 
Therefore, in this paper, we present AgentArcEval, a novel agent architecture evaluation method designed specially to address the complexities of FM-based agent architecture and its evaluation. Moreover, we present a catalogue of agent-specific general scenarios, which serves as a guide for generating concrete scenarios to design and evaluate the agent architecture. We demonstrate the usefulness of AgentArcEval and the catalogue through a case study on the architecture evaluation of a real-world tax copilot, named Luna. 

\end{abstract}

\textbf{Key terms - } Foundation model, large language model, LLM, agent, architecture, evaluation, responsible AI, AI safety.

\section{Introduction}
Foundation models (FMs) are large-scale pretrained models with tens of billions of parameters~\cite{bommasani2021opportunities}, which can be adapted to perform a wide variety of downstream tasks. 
However, FMs exhibit inherent limitations, particularly when facing complex tasks which require users to provide prompts at each step, which can be inefficient and prone to errors. 
An agent is an autonomous system that perceives its environment and takes actions to achieve goals on behalf of humans~\cite{ISO22989_2022}.
The emergence of FMs has enabled the development of highly capable and autonomous agents that can perceive context, reason, plan, and execute workflows to accomplish human goals~\cite{lu2024towards}. 
Given their huge potential to enhance productivity, there has been a surge of interest in FM-based agents across various domains, such as Devin\footnote{https://www.cognition.ai/blog/introducing-devin} and AI Scientist\footnote{https://sakana.ai/ai-scientist}. Throughout this paper, we use the term ``agents'' specifically to refer to FM-based agents.



Software architecture evaluation is a key practice for assessing the extent to which architectural design decisions meet quality attributes and address their tradeoffs~\cite{bass2021software}. Evaluating the architecture of agents is particularly important, as agents are complex compound AI systems that integrate FMs with a variety of out-of-model components, including context engines, prompt optimisers, reasoning and planning, workflow execution, memory, external knowledge bases and tools, guardrails, etc. 
Given these complexities, architectural decisions significantly impact quality attributes of agents, requiring careful balancing of trade-offs. Effective architectural evaluation helps identify potential risks, validate design decisions and improve the agent's architecture to better meet its intended qualities. It also fosters a deeper understanding of how architectural decisions affect the agent's ability to operate, learn and adapt within dynamic environments in a responsible and safe manner.

Over the past decades, various architecture evaluation methods have been developed, such as Architecture Tradeoff Analysis Method (ATAM)~\cite{kazman2000atam}. However, these traditional methods fall short when it comes to evaluating the agent architecture due to their unique characteristics. 
First, goals are special in the context of agents because the primary purpose of using agents is that their users only need to provide high-level goals. The agents need to decompose these high-level user goals into open-ended fine-grained sub-tasks, generate system requirements dynamically, and design and assemble the system implementation at runtime, potentially generating a custom system structure for every user query. Each of the runtime architectural choices made by an agent may have direct implications for quality attributes. As a result, architecture evaluation must move beyond static structure analysis to consider high-level goals, exemplar context scenarios, and guardrails that guide these dynamic runtime trade-offs. The architecture should be assessed for its ability to surface and refine system requirements on the fly - such as goal decomposition, ambiguity resolution, uncertainty handling - rather than its ability to satisfy pre-defined requirements. Consequently, the evaluation focus shifts towards adaptability, observability, and transparency in goal translation, rather than simply functional coverage or satisfaction of stable quality attribute requirements.
Second, agents can autonomously make user requirement-level tradeoff decisions (e.g., cost vs. convenience)  within predefined guardrails, as long as they operate within those boundaries. Architects still need to consider the system requirement-level tradeoffs by designing the guardrail mechanism requirements, such as compute budget caps, trustworthiness scores, ethical constraints, and fall back behaviours. Thus, guardrail design becomes the principal architectural control point, replacing much of the fine-grained pre-specification that traditional architectures had to manage. Third, as agents continuously evolve, the previously assessed risks can change over time or emerge in unexpected ways. Compared to  adaptive systems, agents can retrain, rewire workflows, and adjust external tool selection continuously based on fresh feedback. Also, the agency comes from FM's opaque internals, not explicit rules or well-understood machine learning (e.g., decision tree updates). Thus, runtime assurance mechanisms, behavioural monitoring, and periodic architecture re-evaluation are essential to detect when quality assumptions no longer hold.
Lastly, there is a lack of agent-specific scenarios that capture the inter-artefact interactions, going beyond solely emphasising user-driven quality concerns. 

\begin{figure*}
\centering
\includegraphics[width=0.85\textwidth]{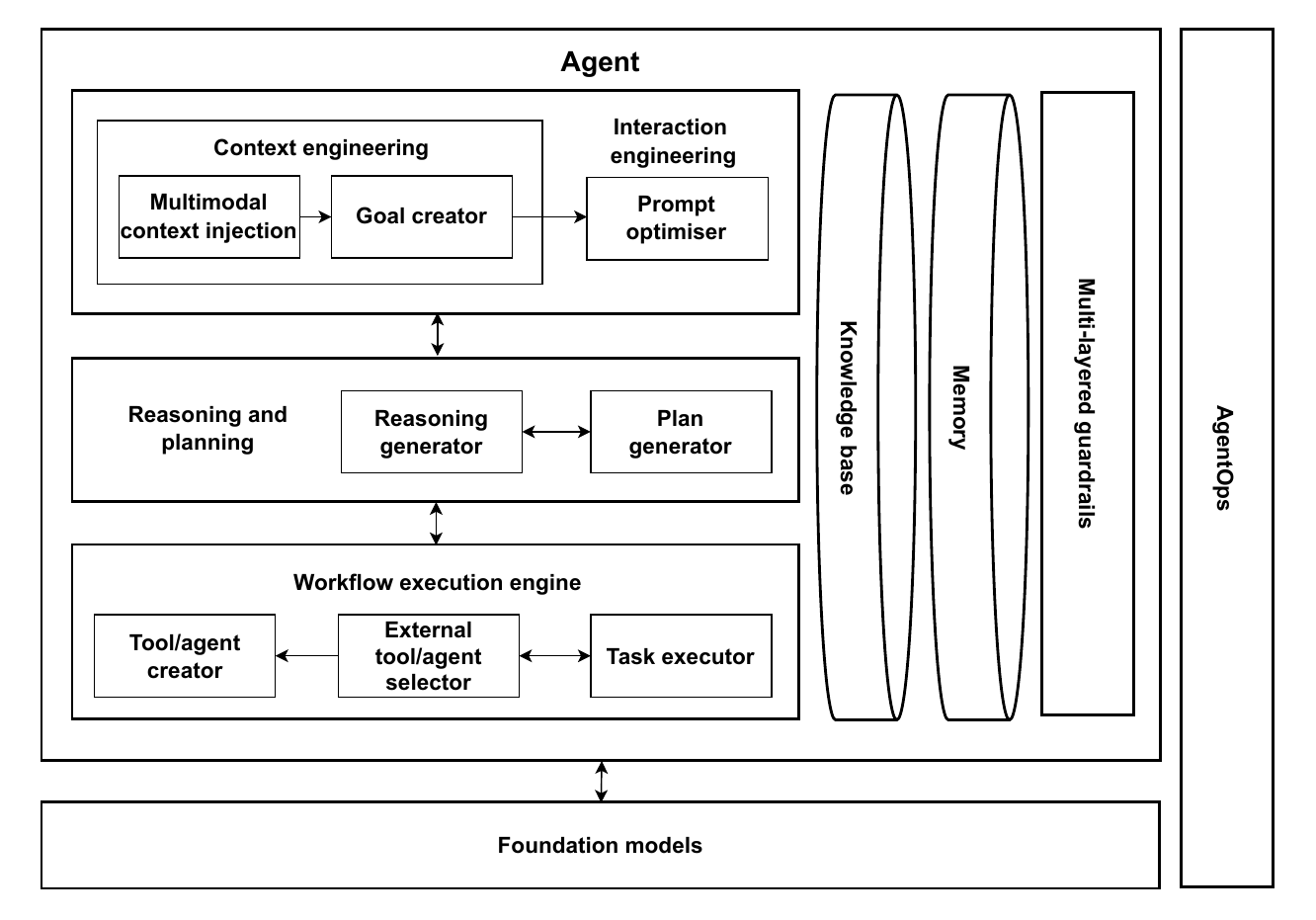}
\caption{Agent Reference Architecture~\cite{lu2024towards}.} \label{fig:agent}
\vspace{-2ex}
\end{figure*}

Therefore, based on the classic ATAM method, we introduce \textbf{AgentArcEval}, an agent architecture evaluation method designed specially to address the complexities of agent architecture and its evaluation. To support the agent architecture evaluation using AgentArcEval, we present a \textbf{catalogue of agent-specific general scenarios}, providing a structured guide for generating concrete scenarios to design and evaluate agent architecture. General scenarios have been shown to have great value in stimulating and guiding the creation of architectural requirements. 

We demonstrate the usefulness of AgentArcEval and the catalogue through a case study evaluating the architecture of the Luna tax copilot\footnote{https://empathetic-ai.com/home}. Please note that the proposed AgentArcEval method and the catalogue of general scenarios in this paper are also applicable to agentic systems, which are designed to prompt an FM multiple times using agent-like design patterns and exhibit varying degrees of agent-like behavior.

The rest of the paper is organised as follows. Section 2 discusses the related work. Section 3 introduces the methodology. Section 4 presents the AgentArcEval method. Section 5 introduces the general scenario catalogue. Section 6 discusses the evaluation. Section 7 concludes the paper.

\section{Background and Related Work}
\subsection{Background}
There has been extensive research on the design of agents. In this work, we build on our previously proposed reference architecture for FM-based agents~\cite{lu2024towards}, which outlines a modular, layered view of agent components and their interactions. Fig.~\ref{fig:agent} illustrates a simplified version of the agent reference architecture.
While goals are typically articulated by users and sent through prompts, some work explores proactive goal anticipation by analysing multimodal context information~\cite{zhao2023seehow,zeng2023gesturegpt}, such as screen recording, eye tracking, and gestures.
 {Agent memory} is structured into short-term memory and long-term memory. Short-term memory contains configurations, recent events, and working context~\cite{packer2023memgpt, jin2023surrealdriver, zhao2024expel, zhang2023building}. The capability of short-term memory is limited by the context window length of the FM.
To extend the storage capacity of  {agent memory}, long-term memory is adopted to store the entire history of processed events as well as accumulated knowledge and past experiences~\cite{packer2023memgpt, jin2023surrealdriver}. 
Achieving a human goal needs reasoning and planning, such as Chain-of-Thought~\cite{wei2022chain} and Tree of Thought~\cite{yao2024tree}. 
Plan reflection allows the agent refine the plan based on feedback~\cite{park2023generative}, which can be done through self-reflection~\cite{yao2022react, madaan2024self, sumers2023cognitive, shinn2023reflexion}, cross-reflection~\cite{chen2023interact, shinn2023reflexion, talebirad2023multi}, or human reflection~\cite{huang2022inner, sarch2023open}. 
The execution engine is responsible for acting the plan, with a task monitor overseeing task status and managing queued tasks~\cite{chen2023autoagents, nascimento2023self}.
The agent may rely on internal FM knowledge or external retrieval augmented generation (RAG) for information, cooperate with other agents or use external tools~\cite{ruan2023tptu, kong2023tptu} to execute the plans. The tool/agent selector~\cite{xie2023openagents, talebirad2023multi} can search registries~\cite{ruan2023tptu, wang2023voyager,kong2023tptu} or on the web to find suitable tools and agents to task completion. 
Guardrails are designed to monitor and control the performance and behaviour of agents. Guardrails can apply to prompts, FMs, reasoning and planning, workflow execution, external tools and knowledge bases~\cite{shamsujjoha2024guardrails}, etc. 

\subsection{Related Work}
 Software architecture evaluation plays a key role in ensuring that architectural decisions align with quality attributes and stakeholder needs. 
Bucaioni et al. recently proposed a high-level functional decomposition for systems that integrate large language models~\cite{bucaioni2025functional}. Their architecture emphasises modularity, interoperability, and quality attributes, and provides a general conceptual structure applicable to a wide range of LLM-based systems. In contrast, our reference architecture~\cite{lu2024towards} targets agentic systems built on foundation models, where components such as context engineering, reasoning and planning, memory, and workflow execution are central. Moreover, our approach is pattern-oriented and explicitly embeds multi-layered guardrails into the architectural layers.
 
 Sobhy et al.~\cite{sobhy2021evaluation} and Fatima and Lago~\cite{fatima2023review}  {are two recent studies performing extensive literature reviews on the software architecture evaluation methods developed over the years.}  {From the perspective of autonomy, architecture evaluation methods can be categorised based on the degree of human involvement. Some methods are entirely reliant on human input, depending on stakeholders to evaluate the design options}~\cite{pooley2010cpasa}. Others adopt a semi-autonomous process, incorporating human-in-the-loop to guide the evaluation~\cite{ghezzi2013dealing}. A distinct subset of methods operate fully autonomously, conducting evaluation without human intervention~\cite{sobhy2020run, van2018cost}.

 In terms of the evaluation stage, various methods cover design-time~\cite{kazman2000atam, nord2003integrating, bengtsson1998scenario, bengtsson2004architecture, eloranta2014lightweight, faniyi2011evaluating}, run-time~\cite{yang2009quality, heaven2009case, moreno2016efficient, esfahani2013learning}, and continuous evaluation~\cite{christensen2010lightweight, pooley2010cpasa, sobhy2020run}.
 Some of the design-time architecture evaluation approaches, such as ATAM~\cite{kazman2000atam} and CBAM~\cite{nord2003integrating}, handle tradeoffs manually through the analysis of tradeoff points elicited from stakeholders, or do not consider it at all (such as SAAM~\cite{kazman1994saam} and ALMA~\cite{lassing2002experiences}). As for the run-time architecture evaluation approaches, some run-time approaches provide automatic management of tradeoffs~\cite{sobhy2020run}, whereas one noticeable investigation is that many approaches have no support for tradeoff management~\cite{yang2009quality}. 

Architecture can experience epistemic uncertainty and/or aleatory uncertainty. Most design-time architecture evaluation methods focus on mitigating epistemic
uncertainty~\cite{bengtsson1998scenario, bengtsson2004architecture, kazman1994saam}. Conversely, aleatory uncertainty is more commonly encountered in run-time architecture evaluation methods~\cite{faniyi2011evaluating, kim2009reinforcement}. A limited number of methods address both epistemic and aleatory uncertainty, including certain design-time methods~\cite{faniyi2011evaluating, ionita2004scenario, meedeniya2011architecture}, run-time methods~\cite{esfahani2013learning, esfahani2011taming}, and continuous methods~\cite{sobhy2020run}.
 Uncertainties and risks are often addressed by developing a set of scenarios that reflect stakeholder envision the system usage~\cite{kazman2000atam, koziolek2011sustainability}. The accuracy and relevance of the evaluation highly depend on the choice of these scenarios. For instance, ATAM identifies and documents risks that could hinder the achievement of quality attribute goals. This includes risks stemming from architectural decisions that negatively impact quality attributes, points of sensitivity where minor changes result in significant quality shifts, and tradeoffs where one decision influences multiple attributes~\cite{jones2001using}. While ATAM focuses on the risks and benefits of architecture decisions, it does not explicitly take into account the cost. CBAM~\cite{nord2003integrating}  extends ATAM by integrating cost-benefit tradeoff analysis into the evaluation process. Scenario-based design-time architecture evaluation methods, such as ATAM~\cite{kazman2000atam},  {CBAM}~\cite{nord2003integrating}, ATMIS~\cite{faniyi2011evaluating}, and APTIA~\cite{kazman2006essential}  partially address uncertainty through the use of scenarios. However, these methods primarily operate at design time and are heavily reliant on stakeholder input. 

 Traditional software architecture evaluations~\cite{babar2004framework, sobhy2021evaluation, fatima2023review}, such as ATAM~\cite{kazman2000atam} and SAAM~\cite{kazman1994saam}, provide detailed insights into evaluation steps and artifacts. However, there is lack of guidance tailored for agent architecture evaluation, especially insufficient consideration of the different ways of building agents and their impact on architecture evaluation.

\section{Methodology}
 To develop the AgentArcEval method and the corresponding catalogue of agent-specific general scenarios, we undertook a structured approach that combined insights from our hands-on project experience with a systematic literature review (SLR) of existing architecture evaluation papers. We began by leveraging our project experience in designing and evaluating the architecture of agents, such as tax copilot, scientific agent, and tender evaluation agent. Through these projects, we identified the unique characteristics of agents, quality attributes essential for agents, frequently encountered scenarios in agent deployment, common design patterns of agents. 

 We then conducted an SLR of prior software architecture evaluation methods by searching the key words ``software architecture evaluation'', ``software architecture assessment'', ``software architecture analysis'' on Google Scholar with time period constraint ``until 2 Jan 2025". The initial search result returned 1080 papers, and after screening, snowballing and selection through a set of criteria, the finalised set included 119 primary studies. Specifically, the inclusion and exclusion criteria are listed as follows.

\textbf{Inclusion criteria:}
 \begin{enumerate}  
     \item A paper that presents a novel software architecture evaluation method/tool.

     \item A paper that presents a case study.
 \end{enumerate}

 \textbf{Exclusion criteria:}
 \begin{enumerate}  
     \item A survey/review paper that analyses existing software architecture evaluation methods.

     \item A paper that proposes analysis frameworks for architecture evaluation methods.

     \item A paper that is not written in English.

     \item Conference version of a study that has an extended journal version.

     \item PhD/Master's dissertations, tutorials, editorials, books.
     
 \end{enumerate}
 
 We examined existing architecture evaluation methods with four main research questions: i) Can existing software architecture evaluation methods be applicable to evaluating the architecture of foundation model based agents? ii) Does the method support autonomy? iii) Does the method support continuous evolution? iv) Does the method provide general scenarios? Through this SLR, we found that they are not appropriate ``out of the box'' for evaluating  agent architectures due to the unique challenges of such systems.~\footnote{Protocol for review purpose only: \url{https://docs.google.com/document/d/1cPuQN5PqPC81VNa8W7ktMFwE6pH-isR4/}}~\footnote{
Extraction sheet for review purpose only: \url{https://docs.google.com/spreadsheets/d/1fHzD7EUT_hdE_76uSlOZ2-Sh_Po7jrwS/}} 
 By combining practical insights from real-world projects with an SLR of existing methods, we were able to design the AgentArcEval method and the general scenario catalogue to address both theoretical gaps and operational considerations of agent architecture evaluations. We used a real-world agentic system to evaluate the usefulness of the proposed method and catalogue.

 In terms of methodology limitations and mitigation strategies, while the AgentArcEval method and scenario catalogue were developed based on insights from a limited number of real-world projects, we have previously performed an SLR on agents~\cite{lu2024towards} and have a thorough understanding of agent design principles and practices. The general scenarios and quality attributes within the catalogue were initially derived from the authors' project experience and architecture expertise. To minimise potential bias, we reviewed the classic software/AI architecture books~\cite{bass2021software,bass25engineering} as well as responsible AI principles~\cite{lu2023responsible}. This allowed us to systematically identify the critical quality attributes essential for the development and operation of AI agents with confidence. Additionally, given the rapid evolution of FMs and agents, there is a risk that the AgentArcEval method or scenario catalogue may become outdated or misaligned with emerging best practices. To address this, we plan to adopt a community-driven approach by publishing the method and catalogue as a living document and inviting contributions from the broader research and development community.

\begin{figure*}
\centering
\includegraphics[width=\textwidth]{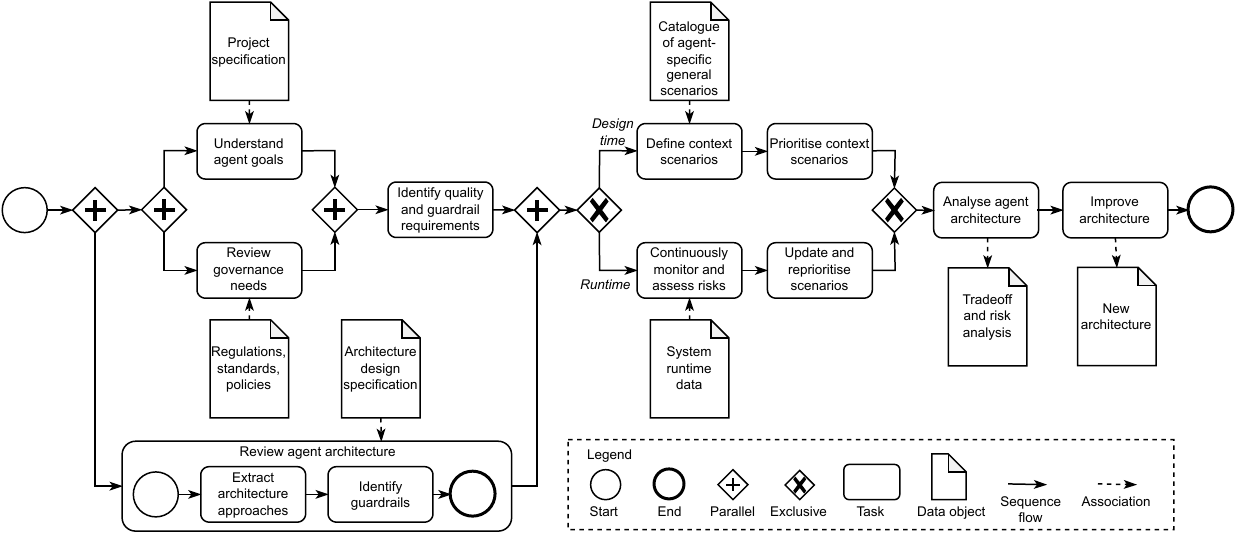}
\caption{The process of AgentArcEval.} \label{fig:process}
\vspace{-2ex}
\end{figure*}

\section{The AgentArcEval Method}

In this section, we present the AgentArcEval method, a scenario-based architecture evaluation method tailored specially for FM-based agents. The method builds on established principles and process from the ATAM, particularly its use of quality attribute scenarios and stakeholder-driven analysis. However, FM-based agents present new architectural challenges, such as goal-directed autonomy and continuous evolution. These aspects are not adequately addressed by traditional evaluation methods and require tailored extensions to ensure meaningful assessment. AgentArcEval addresses this gap by incorporating agent-specific artefacts and guardrails into the evaluation process. The method enables teams to reason explicitly about common decision points in FM-based agent design. It supports early-stage analysis of quality trade-offs through structured, context-specific scenarios, helping stakeholders assess how an architecture responds to real-world operational demands.

\subsection{User types}
The AgentArcEval method can be primarily applied by two types of users, each with distinct objectives.
\begin{itemize}
    \item \textbf{Project team architects}: 
    \begin{itemize}
        \item To assess and refine their own agent architectures to meet project goals and governance needs.
        \item To incorporate mechanisms for runtime architecture evaluation and continuous evolution into their architectures.
    \end{itemize}
    \item \textbf{External architecture experts}: 
    \begin{itemize}
        \item To conduct independent architecture evaluations, with a focus on improving the agent architecture.
        \item To audit the project for quality assurance and regulatory compliance.
    \end{itemize} 
\end{itemize}

\subsection{Evaluation time} 
The AgentArcEval method can be used at various project stages.
\begin{itemize}
    \item \textbf{Early in the project:} Project team architects can apply AgentArcEval after the initial architecture design to evaluate and refine critical design decisions before substantial resources are committed. This ensures that the architecture is aligned with the project goals, while also addressing potential issues/risks early on.
    \item \textbf{Before deployment:} Both project team architects and external architecture experts can apply AgentArcEval towards the end of development, before deployment, to conduct a comprehensive review of the architecture as a check of quality and governance needs.
    \item \textbf{After deployment:} Both project team architects and external architecture experts can use AgentArcEval post-deployment to assess how well the architecture performs in a real-world environment. This can help identify any emerging issues/risks that have not been identified during development and provides opportunities for improvement and evolution.
    \item \textbf{Continuously at runtime:} Although post-deployment architecture evaluation and adaptation is expensive, the AgentArcEval method can be integrated into the agent architecture to enable continuous risk assessment, runtime architecture evaluation and adaptive evolution, ensuring proactive risk management throughout the system's operation.
\end{itemize}

\subsection{Inputs and outputs} 
The following inputs are required for using AgentArcEval.
\begin{itemize}
    \item \textbf{Project specification}: A document outlining the agent goals, intended usage context, and targeted capabilities is valuable for architecture evaluation. Since such documentation is often incomplete or lacks depth in practice, it is important to actively engage with the project team to gather and refine this information ahead of time, or in real time \cite{bass-ieeesw02}.
    \item \textbf{Architecture design specification}: A detailed document describing the overall architecture, including the architecture approaches employed, key design decisions made, and an initial tradeoff and risk analysis. As with the project specification, it is often necessary to interact with the project's leadership in advance to acquire, or generate, this information. 
    \item \textbf{Governance needs}: The relevant AI governance regulations, standards and policies, which the project must comply with. 
        
    \item \textbf{Catalogue of agent-specific general scenarios}: Context scenarios can be defined based on the identified quality attributes and the catalogue of agent-specific general scenarios provided here.
    \item \textbf{System runtime data}: If the system has been deployed, data can be collected during its operation, providing useful insights into its real-world behaviour and performance, which are important for identifying changing or new risks and evaluating the architecture's effectiveness during runtime. 
    \item \textbf{Stakeholder insights}: Inputs from project stakeholders regarding the project's goals, requirements, guardrail constraints, context scenarios, and the prioritisation of different scenarios.
\end{itemize}

The output of AgentArcEvaluation is architecture evaluation results, including \textbf{a tradeoff analysis}, \textbf{a list of identified risks}, and \textbf{recommendations for improving the architecture}.

\subsection{Steps} 
Fig.~\ref{fig:process} illustrates the key steps involved in the AgentArcEval method. To begin with, a user needs to comprehend the goals of a particular agent, review top-level governance requirements, and also the agent architecture design. In particular, the user should identify the quality and guardrail requirements after understand the goals and scrutinising governance needs.

\begin{itemize}
    \item \textbf{Understand agent goals}: The process starts with establishing a clear understanding of the the agent's goals and intended capabilities. When the goals are only vaguely specified, the evaluation team should work closely with stakeholders to elicit, clarify and refine them as part of the architecture evaluation.
    \item \textbf{Review governance needs}: A thorough review of relevant AI governance regulations, standards, and policies is conducted to map to quality/guardrail requirements.
    \item \textbf{Identify quality and guardrail requirements}: The quality/guardrail requirements and associated metrics, which the agent architecture must meet, should be defined.
    \item \textbf{Review agent architecture}: An interactive review of the current agent architecture is performed, examining its components, structure, and interactions. This includes collaboratively identifying architecture approaches used such as key design patterns, tactics, design decisions, and guardrails implemented within the architecture.
\end{itemize}

Subsequently, the user needs to split the architecture evaluation in terms of design time and runtime. The Design time side include defining context scenarios and prioritising context scenarios, while runtime side consists of considering continuous risk monitoring and assessment, as well as updating and reprioritising scenarios according to agent operation. Finally, the user can integrate the results to analyse and improve the agent architecture. Detailed process of each step is listed as follows. 

\begin{itemize}
    \item \textbf{Define context scenarios}: Based on the identified quality and guardrail requirements, context scenarios are developed to explore the various operational environments and use cases in which the agent will function. These scenarios help evaluate how well the architecture supports the required qualities under realistic conditions.
    \item \textbf{Prioritise context scenarios}: The context scenarios are ranked based on their importance, potential impact and risk, and relevance to agent goals. 
    \item \textbf{Analyse agent architecture}: A thorough analysis of the architecture is conducted to evaluate how well it supports the agent’s goals, meets quality requirements, and complies with guardrails and regulations. This analysis is guided by available artifacts such as design documents, architecture diagrams, system logs, evaluation results, and structured tools like checklists or assessment templates. Key architectural tradeoffs and risks are identified.
    \item \textbf{Improve architecture}: Based on the analysis, recommendations for improving the architecture are made to better meet the agent's goals and mitigate identified risks. This may involve refining design decisions, adding or changing architecture approaches.
    \item \textbf{Continuously monitor and assess risks}: Continuous monitoring and assessment are performed to detect any emerging or evolving risks during runtime.
    \item \textbf{Update and reprioritise scenarios}: As the agent evolves, it is important to periodically revisit and update the context scenarios and their prioritisation. This triggers a runtime evaluation of the architecture by revisiting key steps including \textbf{analysing agent architecture} and \textbf{improving architecture} to ensure the architecture can meet changing usage contexts and risks. 
\end{itemize}

\section{General Scenarios}

In this section, we present a catalogue of agent-specific general scenarios for quality attributes including accuracy, adaptability, efficiency, privacy, security, fairness, availability, observability, transparency, safety, and contestability. In particular, the quality attributes were carefully selected as they are regarded critical to the development and operation of AI systems~\cite{bass25engineering, bass2021software} and their alignment with responsible AI principles~\cite{lu2023responsible}. They present a starting point derived from both literature review and practical considerations in agent systems.
We constructed the catalogue by first extracting a long-list of situations from three deployed agent projects and a systematic literature review, then retaining one exemplar scenario for each of the eleven quality attributes, including accuracy, adaptability, efficiency, privacy, security, fairness, availability, observability, transparency, safety, and contestability. A scenario entered the baseline catalogue only when it met two checks: (1) the situation has occurred in an actual FM-agent deployment, guaranteeing practical relevance; (2) it can be written in the standard ``Source → Stimulus → Environment → Artefact → Response → Measure" format, so its outcome can be verified objectively. This screening produces a lean but comprehensive starter set, which teams can further extend or refine during the ``Define Context Scenarios" workshop to address emerging, domain-specific concerns. As the field evolves, new quality attributes and corresponding scenarios will inevitably surface; the eleven scenarios presented here are intended only as a foundational baseline for evaluation.


Each scenario is described using the traditional scenario template~\cite{bass2021software}:
\begin{itemize}
    \item Source: The entity (e.g., user or external auditor) or artifact (e.g., runtime evaluator) that generated the stimulus.
    \item Stimulus: The event or condition that requires a response from the system.
    \item Environment: A context where the stimulus occurs.
    \item Artefact: The agent artefacts that are stimulated.
    \item Response: The activity undertaken as a result of the arrival of the stimulus.
    \item Response measure: The metrics used to evaluate whether the response is satisfactory. The sources of response measures are from~\cite{bass25engineering, bass2021software}.
\end{itemize}

\begin{figure}
\centering
\includegraphics[width=0.64\columnwidth]{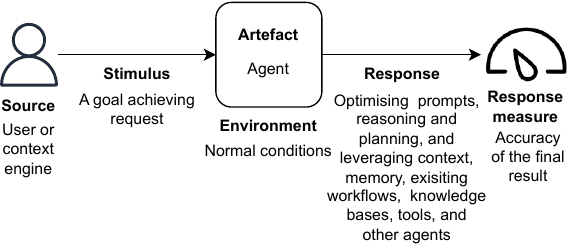}
\caption{Accuracy general scenario.} \label{fig:accuracy}
\vspace{-2ex}
\end{figure}

\subsection{Accuracy General Scenario}
Fig.~\ref{fig:accuracy} shows the portions of accuracy general scenario.
\begin{itemize}
    \item \textbf{Source}: A user relying on the agent to achieve goals, or context engine proactively suggesting goals to the user.
    \item \textbf{Stimulus}: A goal achieving request initiated by the user or proactively suggested by the agent's context engine.
    \item \textbf{Artefact}: Agent.
    \item \textbf{Environment}: The agent operates in normal conditions.
    \item \textbf{Response}: The agent accurately accomplishes the goal through optimised prompts, reasoning and planning, past agent memory, existing workflows, knowledge bases, external tools, and other agents as needed.
    \item \textbf{Response Measure}: The accuracy of the final result in relation to the user expectation.
\end{itemize}

\begin{figure}
\centering
\includegraphics[width=0.63\columnwidth]{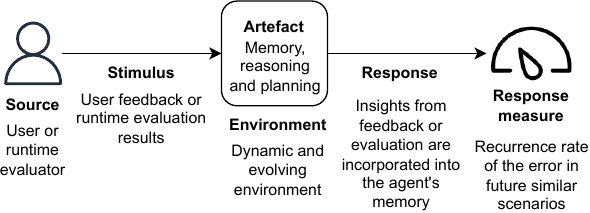}
\caption{Adaptability general scenario.} \label{fig:adaptability}
\vspace{-2ex}
\end{figure}

\subsection{Adaptability General Scenario}
Fig.~\ref{fig:adaptability} shows the portions of adaptability general scenario.

\begin{itemize}
    \item \textbf{Source}: A user who provides feedback on the agent's outputs (either final or intermediate ones), or an automated runtime evaluator that continuously monitors and evaluates these outputs.
    
    \item \textbf{Stimulus}: The agent receives feedback from the user, or it gets evaluation results from the runtime evaluator.
    
    \item \textbf{Artefact}: Agent memory, reasoning and planning.
    \item \textbf{Environment}: A dynamic environment where the agent continuously collects and processes user feedback and runtime evaluation results.

    \item \textbf{Response}: The agent updates its memory to incorporate the feedback or evaluation insights. The updated memory can adapt the agent's future prompts, reasoning and planning, and workflow execution.

    \item \textbf{Response Measure}:  Rate of successful adaptations, time to adapt.
\end{itemize}

\begin{figure}
\centering
\includegraphics[width=0.63\columnwidth]{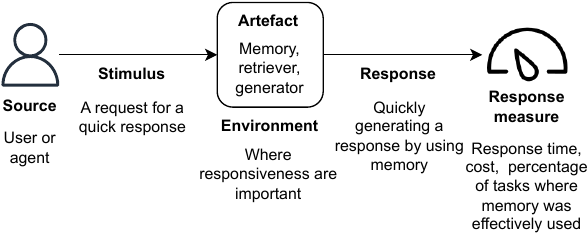}
\caption{Efficiency general scenario.} \label{fig:efficiency}
\vspace{-2ex}
\end{figure}

\subsection{Efficiency General Scenario}
Fig.~\ref{fig:efficiency} shows the portions of efficiency general scenario.

\begin{itemize}
    \item \textbf{Source}: A user or the agent itself triggering the need for a quick response based on previous interactions or learned experiences stored in the agent memory.
    \item \textbf{Stimulus}: A request where the agent can utilise relevant memory from past interactions or task executions to quickly generate a response, minimising the need for redundant processing or reasoning.
    \item \textbf{Artefact}: Agent memory, retriever, generator.
    \item \textbf{Environment}: An environment where responsiveness is important.
    \item \textbf{Response}: The agent quickly generates a response by using relevant agent memory, such as previous results, existing workflows, or past experience.
    \item \textbf{Response Measure}: The response time, cost.
\end{itemize}

\begin{figure}
\centering
\includegraphics[width=0.65\columnwidth]{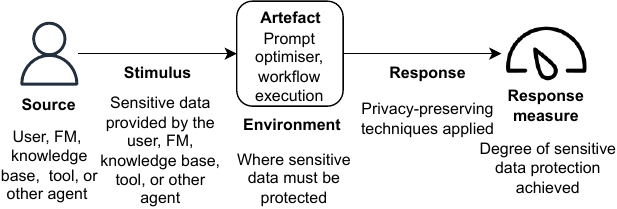}
\caption{Privacy general scenario.} \label{fig:privacy}
\vspace{-2ex}
\end{figure}

\subsection{Privacy General Scenario}
Fig.~\ref{fig:privacy} shows the portions of privacy general scenario.

\begin{itemize}
    \item \textbf{Source}: A user, FM, knowledge base, external tool, or other agent providing sensitive data.
    \item \textbf{Stimulus}: Sensitive data provided by the user, FM, knowledge base, external tool, or other agent.
    \item \textbf{Artefact}: Workflow execution.
    \item \textbf{Environment}: An environment where sensitive data must be protected.
    \item \textbf{Response}: Privacy-preserving techniques, such as data anonymisation, encryption, and differential privacy, are applied to protect sensitive data. 
    \item \textbf{Response Measure}: Percentage of protected sensitive data.
\end{itemize}

\begin{figure}
\centering
\includegraphics[width=0.65\columnwidth]{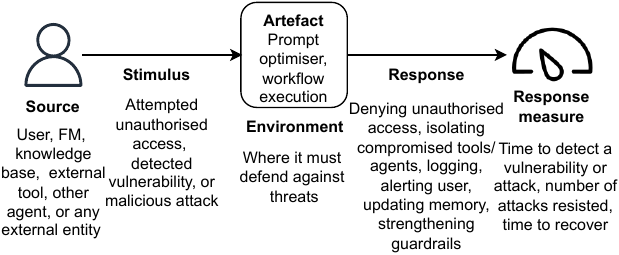}
\caption{Security general scenario.} \label{fig:security}
\vspace{-2ex}
\end{figure}

\subsection{Security General Scenario}
Fig.~\ref{fig:security} shows the portions of security general scenario.

\begin{itemize}
    \item \textbf{Source}: A user, FM, external tool, other agent, or any external entity attempting unauthorised access or posing potential vulnerabilities or threats.
    \item \textbf{Stimulus}: An attempted unauthorised access, a detected vulnerability, a malicious attack.
    \item \textbf{Artefact}: Prompt optimiser, workflow execution.
    \item \textbf{Environment}: A potentially hostile environment, where it must defend against a variety of threats while maintaining normal operations and overall system integrity.
    \item \textbf{Response}: The agent detects the security threats and takes immediate actions to mitigate them. This includes analysing  vulnerabilities, denying unauthorised access, isolating compromised tools or agents, logging incidents, alerting the user or administrator, updating agent memory, and strengthening guardrails.
    \item \textbf{Response Measure}: Time to detect a vulnerability, time to recover from a successful attack.
\end{itemize}

\begin{figure}
\centering
\includegraphics[width=0.6\columnwidth]{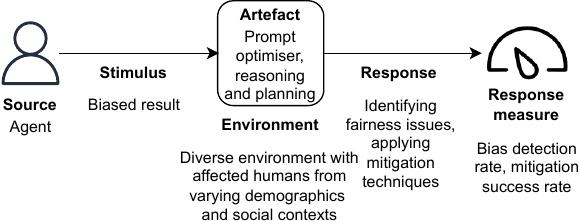}
\caption{Fairness general scenario.} \label{fig:fairness}
\vspace{-2ex}
\end{figure}

\subsection{Fairness General Scenario}
Fig.~\ref{fig:fairness} shows the portions of fairness general scenario.

\begin{itemize}
    \item \textbf{Source}: A user generating a biased result.
    \item \textbf{Stimulus}:  Bias in the agent's final result, or intermediate result or user.
    \item \textbf{Artefact}: Agent.
    \item \textbf{Environment}: A diverse environment with affected humans from varying demographics and social contexts.
    \item \textbf{Response}: The agent identifies fairness issues and applies mitigation techniques to ensure equitable outcomes.
    \item \textbf{Response Measure}: The bias detection rate, mitigation success rate.
\end{itemize}

\begin{figure}
\centering
\includegraphics[width=0.6\columnwidth]{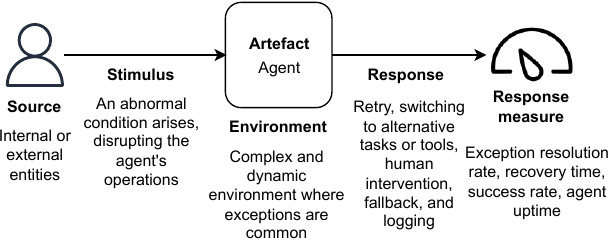}
\caption{Availability general scenario.} \label{fig:reliability}
\vspace{-2ex}
\end{figure}

\subsection{Availability General Scenario}
Fig.~\ref{fig:reliability} shows the portions of availability general scenario.

\begin{itemize}
    \item \textbf{Source}: Internal or external entities that triggers abnormal conditions interrupting the agent's operations.
    \item \textbf{Stimulus}: An abnormal condition arises, disrupting the agent's operations, such as unmet task dependencies, tool failures, data quality issues, or user interventions.
    \item \textbf{Artefact}: Agent.
    \item \textbf{Environment}: A complex and dynamic environment where exceptions are common in task execution, external tool API calls, data operations, and interactions with other agents.
    \item \textbf{Response}: The agent detects the exceptions and activates appropriate handling mechanisms, including retrying failed tasks, switching to alternative tasks or tools, requesting human intervention, applying fallback, using alternative resources, and logging the incident.
    \item \textbf{Response Measure}: The exception resolution rate, mean recovery time, task success rate, agent uptime.
\end{itemize}

\begin{figure}
\centering
\includegraphics[width=0.6\columnwidth]{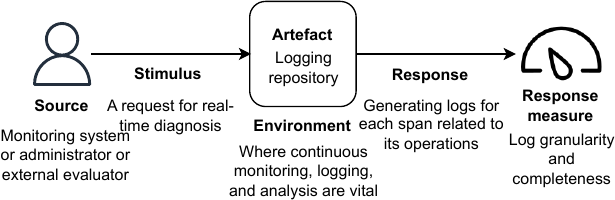}
\caption{Observability general scenario.} \label{fig:observability}
\vspace{-2ex}
\end{figure}

\subsection{Observability General Scenario}
Fig.~\ref{fig:observability} shows the portions of observability general scenario.

\begin{itemize}
    \item \textbf{Source}: A monitoring system, system administrator, or external evaluator, aiming to assess the agent’s performance and behavior in real-time.
    \item \textbf{Stimulus}: A request for real-time diagnosis, triggered by events like performance degradation, abnormal execution.
    \item \textbf{Artefact}: Log repository.
    \item \textbf{Environment}: A complex, real-time environment where continuous monitoring, logging, and analysis are vital.
    \item \textbf{Response}: The agent generates logs for each span, a traceable unit of operation, related to its behaviour, including reasoning span, planning span, workflow span, task span, tool span, evaluation span, and FM span. This provides stakeholders with real-time insights into the agent's health, goal completion progress, and potential anomalies.
    \item \textbf{Response Measure}: Log granularity and completeness.
\end{itemize}

\begin{figure}
\centering
\includegraphics[width=0.6\columnwidth]{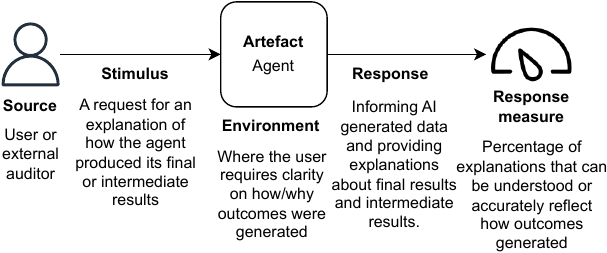}
\caption{Transparency general scenario.} \label{fig:transparency}
\vspace{-2ex}
\end{figure}

\subsection{Transparency General Scenario}
Fig.~\ref{fig:transparency} shows the portions of transparency general scenario.

\begin{itemize}
    \item \textbf{Source}: A user or external auditor seeking to understand the reasoning behind the agent’s final/intermediate results.
    \item \textbf{Stimulus}:  A request for an explanation of how the agent produced its final or intermediate results.
    \item \textbf{Artefact}: Agent.
    \item \textbf{Environment}: 
    A context where the user requires clarity on how/why outcomes were generated.
    \item \textbf{Response}: 
    The agent informs that the outcome was generated by an AI agent and provides a human-understandable explanation of its final and intermediate results. This explanation includes justifications for goal understanding, planning,  {agent memory} and knowledge retrieval, external tool and agent selection, and workflow execution.
    \item \textbf{Response Measure}: The percentage of users who can understand the explanations, the time taken to generate and deliver explanations to the user.
\end{itemize}

\begin{figure}
\centering
\includegraphics[width=0.6\columnwidth]{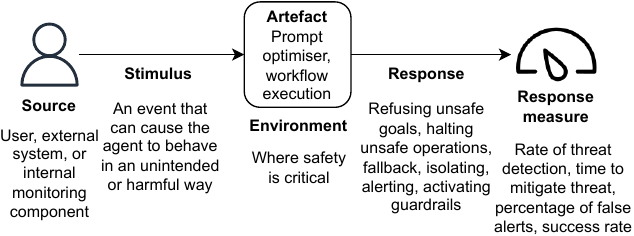}
\caption{Safety general scenario.} \label{fig:safety}
\vspace{-2ex}
\end{figure}

\subsection{Safety General Scenario}
Fig.~\ref{fig:safety} shows the portions of safety general scenario.

\begin{itemize}
    \item \textbf{Source}: A user, external system, or internal monitoring component detecting potential unsafe behaviour.
    \item \textbf{Stimulus}: An event that can cause the agent to behave in an unintended or harmful way.
    \item \textbf{Artefact}: Prompt optimiser, workflow execution.
    \item \textbf{Environment}: 
    An environment where safety is critical.
    \item \textbf{Response}: 
    The agent detects potential safety threats and immediately triggers predefined safety mechanisms, including refusing unsafe human goals, halting unsafe operations, switching to a safe fallback mode, isolating faulty components, alerting humans, activating guardrails.
    \item \textbf{Response Measure}: The rate of threat detection, percentage of unsafe events avoided, percentage of events that require human intervention, the time taken to mitigate a safety threat.
\end{itemize}

\begin{figure}
\centering
\includegraphics[width=0.6\columnwidth]{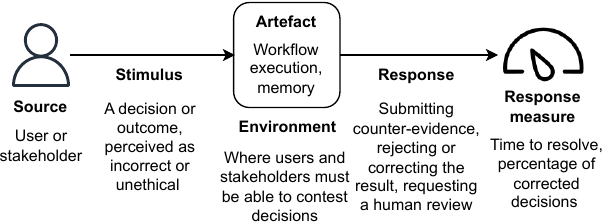}
\caption{Contestability general scenario.} \label{fig:contestability}
\vspace{-2ex}
\end{figure}

\subsection{Contestability General Scenario}
Fig.~\ref{fig:contestability} shows the portions of contestability general scenario.

\begin{itemize}
    \item \textbf{Source}: A user or stakeholder who wants to question or contest the agent’s decision or outcome.
    \item \textbf{Stimulus}: A specific decision or outcome generated by the agent, perceived as incorrect or unethical, which requires further justification or review.
    \item \textbf{Artefact}: Workflow execution, agent memory.
    \item \textbf{Environment}: 
    An environment where users and stakeholders must be able to contest decisions, especially in high-stakes environments such as healthcare, legal, financial, or public-sector applications.
    \item \textbf{Response}: 
   The agent provides a detailed explanation of how outcome was produced. Users challenge the outcome by submitting counter-evidence, rejecting or correcting the result, requesting a human review. 
    \item \textbf{Response Measure}: The time taken to resolve contested decisions, the percentage of contested decisions that result in a meaningful correction after the review process.
\end{itemize}

\section{Case Study}
In this section, we evaluate the usefulness of AgentArcEval and the catalogue of agent-specific general scenarios through a case study on the architecture evaluation of the Luna tax copilot\footnote{https://empathetic-ai.com}. 
The goal is to demonstrate the method's practical applicability in a real-world, operational context. We provide a structured analysis of the architecture evaluation process, demonstrating how the AgentArcEval methodology and the general scenario catalogue guide architectural analysis and decision making.

The evaluation followed a qualitative, scenario-based analysis, following the AgentArcEval process. We first defined a set of context-specific quality scenarios, derived from our general scenario catalogue and refined in collaboration with key stakeholders from the Luna project. These stakeholders include software architects, developers, and product managers, who contributed domain knowledge, technical insights, and operational constraints. Their input helped prioritise quality attributes, validate the realism of the scenarios, and assess how the proposed architectural responses aligned with actual system goals.

Each scenario focused on a critical quality attribute and was written in the standard stimulus-response-measure format. We then assessed how the revised architecture satisfied each scenario, highlighting relevant trade-offs and identifying risk areas where guardrails were necessary. This structured walkthrough helped surface several architectural tensions early in the design process and guided concrete design decisions.

\subsection{Understand agent goals}
After discussions with the project team, the architecture analysts gained a clear understanding that the Luna tax copilot is an agent system to provide professional and accurate tax recommendation, with clear references to the Australian Taxation Office database\footnote{https://www.ato.gov.au/single-page-applications/legaldatabase}, specially for tax professionals.

\subsection{Review governance needs}
The architecture analysts reviewed the Australian Voluntary AI Safety Standard\footnote{https://www.industry.gov.au/publications/voluntary-ai-safety-standard} and identified the following relevant guardrails.
\begin{itemize}
    \item \textit{Protect AI systems, and implement data governance measures to manage data quality and provenance.}
    \item \textit{Test AI models and systems to evaluate model performance and monitor the system once deployed.}
    \item \textit{Enable human control or intervention in an AI system to achieve meaningful human oversight.}
    \item \textit{Inform end-users regarding AI-enabled decisions, interactions with AI and AI-generated content.}
    \item \textit{Establish processes for people impacted by AI systems to challenge use or outcomes.}
    \item \textit{Be transparent with other organisations across the AI supply chain about data, models and systems to help them effectively address risks.}
    \item \textit{Keep and maintain records to allow third parties to assess compliance with guardrails.}
\end{itemize}

\subsection{Identify quality and guardrail requirements}
Based on an understanding of the project goals, deployment context, and relevant AI safety standards, we worked with Luna project stakeholders to identify  seven key quality attributes and guardrail types most critical to the system. While our general catalogue includes eleven, only those deemed directly relevant to the Luna tax copilot's operational risks and design priorities were selected.
\begin{itemize}
    \item Accuracy: Providing correct and precise tax recommendation to tax professionals, ensuring it is aligned with the most up-to-date data from Australian Taxation Office legal database's latest data.
    \item Adaptability: Incorporating user feedback into the  {agent memory} to improve future recommendation generation.
    \item Efficiency: Quickly delivering tax recommendation for queries that are similar to previously answered ones. 
    \item Transparency: Providing clear practical explanations for each recommendation, detailing the copilot's reasoning process and citing the specific legal references and foundation model version were used.
    \item Observability: Enabling stakeholders to track historical queries, monitor user feedback, and receive alerts on copilot health, performance or users complaints.
    \item Contestability: Allowing users to challenge the copilot's recommendation and submit feedback for review.
    \item Privacy: Ensuring all input and output data is desensitised to protect personally identifiable information and sensitive company data.
\end{itemize}

\subsection{Define and Prioritise Context Scenarios}
The architecture analysts defined and prioritised context scenarios based on the goals, identified quality requirements, and the general scenario catalogue. These scenarios and their priorities were then confirmed by the project team. The sources of the response measures are from the requirements of the project and governance needs. Below are the context scenarios for the identified quality attributes that are critical to the tax copilot.

\textbf{Accuracy Context Scenario}
\begin{itemize}
    \item  \textbf{Scenario ID}: 1
    \item  \textbf{Quality Attribute}: Accuracy
    \item  \textbf{Priority}: High
    \item  \textbf{Source}: Tax professional.
    \item  \textbf{Stimulus}: The tax professional submits a query about capital gains tax on the sale of a rental property.
    \item  \textbf{Artefact}: Context engine, reasoning and planning, workflow execution, retriever, generator, vector database.
    \item  \textbf{Environment}: The tax professional uses the copilot during a tax season where accurate recommendation is key.
    \item  \textbf{Response}: The copilot retrieves the latest relevant tax regulations and provides accurate, context-specific tax recommendation.
    \item  \textbf{Response Measure}: The copilot should achieve at least 95\% accuracy in providing relevant responses and using 95\% correct references to the applicable regulations.
\end{itemize}

\textbf{Adaptability Context Scenario}
\begin{itemize}
    \item  \textbf{Scenario ID}: 2
    \item  \textbf{Quality Attribute}: Adaptability
    \item  \textbf{Priority}: Medium
    \item  \textbf{Source}: Tax professional or runtime evaluator component
    \item  \textbf{Stimulus}: A mistake is identified in the system’s recommendation on capital gains tax, either flagged manually by a tax professional or automatically detected by a runtime evaluation component.
    \item  \textbf{Environment}: The feedback is submitted by the tax professional during live system operation via a structured correction interface, or auto-generated evaluation report.
    \item  \textbf{Artefact}:  Agent memory, reasoning and planning
    \item  \textbf{Response}: The system updates its agent memory with the new information and adjusts its reasoning logic to avoid repeating the same error in similar future cases.
    \item  \textbf{Response Measure}: 99\% of valid feedback instances result in a correct update.
\end{itemize}

\textbf{Efficiency Context Scenario}
\begin{itemize}
    \item  \textbf{Scenario ID}: 3
    \item  \textbf{Quality Attribute}: Efficiency
    \item  \textbf{Priority}: Medium
    \item  \textbf{Source}: Tax professional
    \item  \textbf{Stimulus}: A query is submitted about vehicle tax deductions, which has been previously answered.
    \item  \textbf{Environment}: The query is made during peak filling season, requiring a quick response.
    \item  \textbf{Artefact}:  {Agent memory}, retriever, generator
    \item  \textbf{Response}: The tax copilot retrieves the previously provided recommendation on vehicle tax deduction from  {agent memory} and delivers it promptly to the user.
    \item  \textbf{Response Measure}: The response time must be under 1 second for queries previously answered.
\end{itemize}

\begin{figure*}
\centering
\includegraphics[width=\textwidth]{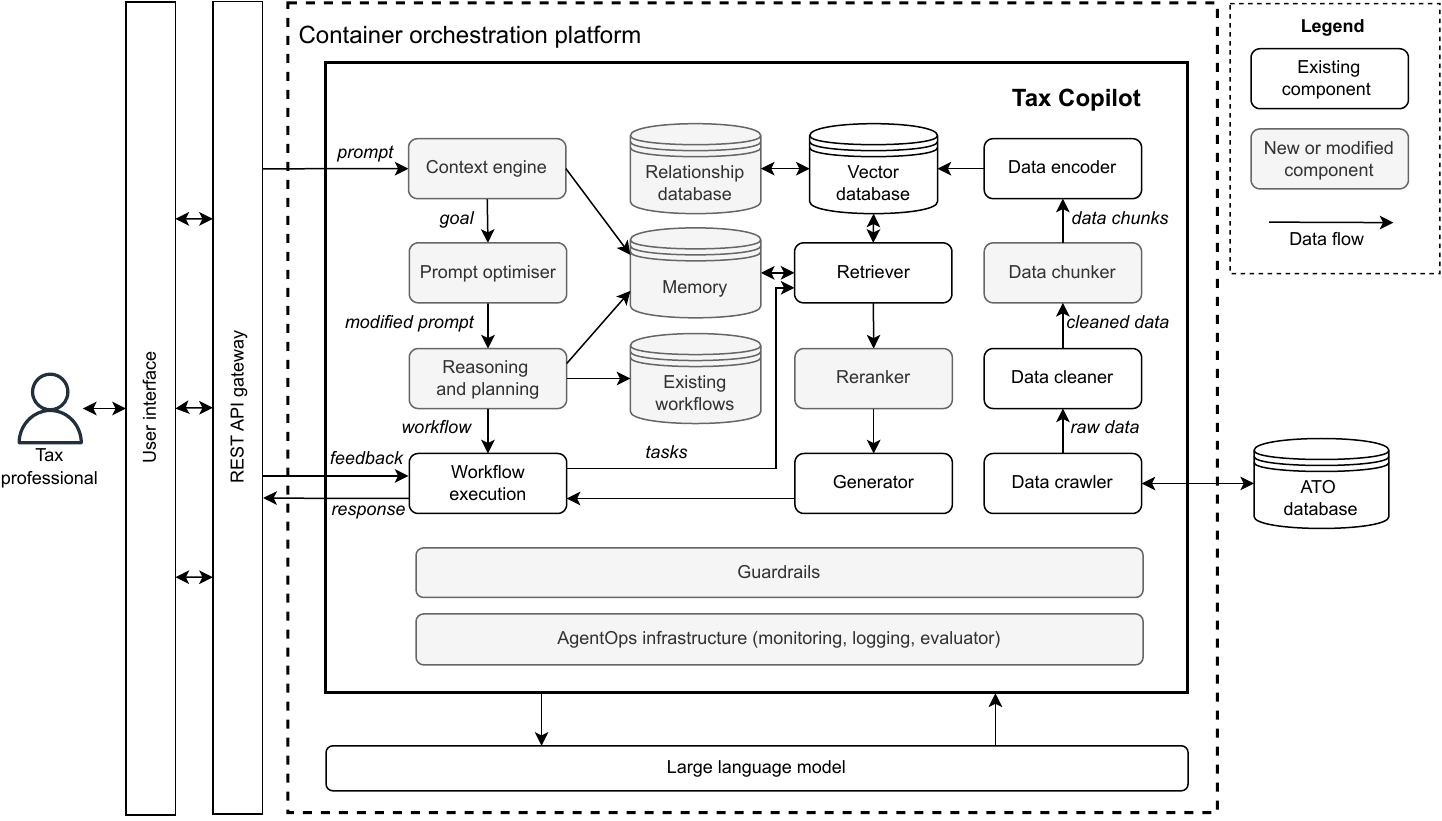}
\caption{Component View of the Luna Tax Copilot Architecture.} \label{fig:luna}
\vspace{-2ex}
\end{figure*}

\textbf{Transparency Context Scenario}
\begin{itemize}
    \item  \textbf{Scenario ID}: 4
    \item  \textbf{Quality Attribute}: Transparency
    \item  \textbf{Priority}: High
    \item  \textbf{Source}: Tax professional.
    \item  \textbf{Stimulus}: The tax professional requests an explanation of how the copilot arrived at its recommendation for income tax deduction.
    \item  \textbf{Environment}: The query is made while preparing tax recommendation reports for a client.
    \item  \textbf{Artefact}: Copilot.
    \item  \textbf{Response}: The tax copilot provides its reasoning process, listing the legal sources consulted and explaining the FM version used\footnote{Recording the FM version enables users, developers, and auditors to verify how a specific response was generated, thereby supporting both traceability and reproducibility. If an error or bias is identified, knowing the exact FM version helps determine whether the issue stems from the model's inherent limitations or from other factors such as prompt design, retrieval context, or system configuration. In our case study, the Luna tax copilot employed OpenAI GPT-4o.}.
    \item  \textbf{Response Measure}: At least 95\% of users understand the explanation, at least 95\% of the references are correct.
\end{itemize}

\textbf{Observability Context Scenario}
\begin{itemize}
    \item  \textbf{Scenario ID}: 5
    \item  \textbf{Quality Attribute}: Observability
    \item  \textbf{Priority}: Medium
    \item  \textbf{Source}: Monitoring component.
    \item  \textbf{Stimulus}:  {An alert is triggered due to an increasing number of low scores provided by the copilot user about recommendation on small business tax deductions.}
    \item  \textbf{Environment}: The copilot is continuously monitored.
    \item  \textbf{Artefact}: Log repository.
    \item  \textbf{Response}: The tax copilot automatically logs all queries, responses, and user feedback.
    \item  \textbf{Response Measure}: Log 100\% queries, responses, and feedback.
\end{itemize}

\textbf{Contestability Context Scenario}
\begin{itemize}
    \item  \textbf{Scenario ID}: 6
    \item  \textbf{Quality Attribute}: Contestability
    \item  \textbf{Priority}: High
    \item  \textbf{Source}: Tax professional
    \item  \textbf{Stimulus}: A tax professional disagrees with the recommendation provided on superannuation contributions and submits feedback to challenge the recommendation.
    \item  \textbf{Environment}: An environment where tax professionals must be able to contest and review the recommendation provided.
    \item  \textbf{Artefact}: Retriever, generator,  {agent memory}.
    \item  \textbf{Response}: The tax copilot logs the feedback and flags the case for review by tax experts.
    \item  \textbf{Response Measure}: 100\% of contested cases are reviewed and resolved within 48 hours.
\end{itemize}

\textbf{Privacy Context Scenario}
\begin{itemize}
    \item  \textbf{Scenario ID}: 7
    \item  \textbf{Quality Attribute}: Privacy
    \item  \textbf{Priority}: High
    \item  \textbf{Source}: Tax professional submitting sensitive data or the generator component producing sensitive data.
    \item  \textbf{Stimulus}: The tax professional submits personal data for tax calculation, or the generator component produces responses that contain sensitive information.
    \item  \textbf{Environment}: The copilot is operated in an environment where sensitive data must be protected.
    \item  \textbf{Artefact}: Prompt optimiser, generator.
    \item  \textbf{Response}: The tax copilot desensitised all sensitive data before processing request or generating responses.
    \item  \textbf{Response Measure}: 99\% of sensitive data must be filtered.
\end{itemize}

\subsection{Analyse and Improve Architecture}

Fig.~\ref{fig:luna} illustrates one view of the architecture of the Luna tax copilot.
The white boxes represent the original architecture before the review while the grey boxes indicate the components modified or added after the review. 
In this section, we will analyse the architecture by mapping to the defined context scenarios.

The proposed architecture components are deployed using containerization technology with Docker, orchestrated via Kubernetes, ensuring scalability and resilience across various workloads \cite{burns2016kubernetes}. Data components such as the Vector Database and Relationship Database are hosted on cloud-managed services (e.g., AWS RDS and Pinecone), enabling efficient scaling, reliable storage, and low-latency access \cite{aws_rds, pinecone_cloud}. Communication between internal components, including the Retriever, Reranker, and Generator, is handled via RESTful APIs using HTTP and JSON payloads, providing structured and efficient data transfer. Data ingestion components such as the Data Crawler fetch information from the ATO database through secure, encrypted channels to ensure compliance and security. All data flow within the system utilizes secure internal networking infrastructure provided by the cloud environment (e.g., AWS Virtual Private Cloud) to maintain data integrity and security standards \cite{aws_vpc}.

To clarify the relationship between the Luna Tax Copilot architecture shown in Fig.~\ref{fig:luna} and the reference architecture shown in Fig.~\ref{fig:agent}, we provide a component-level mapping. The \textbf{context engine} in the Luna architecture directly corresponds to the \textbf{context engineering} component in the reference architecture. Both architectures include a \textbf{prompt optimiser} component and a \textbf{reasoning and planning}. Luna’s \textbf{workflow execution} component is an instantiation of the \textbf{workflow execution engine} in the reference architecture. Similarly, both architectures include a \textbf{memory} component. In Luna, there is a combination of a \textbf{relational database} and a \textbf{vector database}, which together align with the \textbf{knowledge base} in the reference architecture. Luna also incorporates several data pre-processing modules, which can be considered auxiliary components operating either externally or internally, and are therefore not explicitly represented in the reference architecture. RAG components including Retriever, Re-ranker, and Generator functions collectively form part of the \textbf{Task Execution} process in the reference architecture. Furthermore, both architectures feature \textbf{Guardrails} and \textbf{AgentOps} layers, and Luna’s \textbf{Large Language Model} corresponds to the \textbf{Foundation Model} in the reference architecture. This mapping shows that the Luna Tax Copilot architecture concretely instantiates the main functional components and layers of the reference architecture.

\subsubsection{Architecture Analysis for Accuracy Context Scenario}
To ensure accuracy, several architecture approaches have been applied to the previous architecture:
\begin{itemize}
    \item \textbf{Context engine}: A topic classifier has been integrated into the system to categorise user queries before generating responses, allowing the classified topic to serve as contextual input for subsequent processing.
    \item \textbf{Prompt optimiser}: A prompt template is used to help large language model (LLM) understand the queries.
    \item \textbf{Generator}: Regarding the \textbf{response optimisation strategy}, the previous architecture design has included references for each generated piece of recommendation, such as IDs and links to rulings. Response templates have been employed to structure the recommendation, making it easier for tax professional to read and understand the reasoning behind the generated response.
    \item \textbf{Retriever}: The \textbf{retriever} fetches two types of information: private ruling and legislation. Legislation defines general rules, while private rulings cover the practical application of relevant provisions. If private ruling is successfully retrieved, the \textbf{generator} uses the template that combines both private ruling and legislation. If only legislation is retrieved, the legislation-specific template is applied.
    \item \textbf{Data crawler}: Legislation was crawled once since there was no updates after crawling, whilst for private rulings, the team decided to crawl ruling data from the ATO database on a weekly basis instead of daily regarding \textbf{crawling frequency}, to balance the trade-off between accuracy and cost. 
\end{itemize}

In addition to the above, there are other approaches to further improve accuracy. 
\begin{itemize}
    \item \textbf{Context engine}: \textbf{Context definition} is a key design decision for better understanding the context. Beyond query and scenario classification, creating user profile is also helpful. The profile can either be created by having users input their information, or the copilot can continuously learn and update the profile based on interaction. To optimise accuracy, we opted to have users initially create their profile, with the copilot continuing to learn and update the profile based on the subsequent interactions.
    \item \textbf{Prompt optimiser}: Regarding \textbf{prompt optimisation strategy}, we considered generating multiple prompts to improve understanding but decided against the strategy due to the high cost associated with additional LLM queries. Instead, we will continue using the prompt template.
    \item \textbf{Reasoning and planning}: On the \textbf{planning strategy}, we chose to first break down the query and then search relevant legislation followed by relevant rulings. We also discussed creating a repository of \textbf{existing workflows} to potentially reuse of workflows from different companies. However, this is a lower priority, as collecting such workflows maybe difficult due to many companies treating them as proprietary intellectual properties.
    \item \textbf{Data chunker}: In the previous architecture, each crawled private ruling is indexed by date and assigned a unique identifier that maps to a dedicated webpage, and the content of this entire webpage is embedded into vector database. In this circumstance, embeddings can contain significant extraneous data beyond a specific question or subject matter, which may dilute the relevance score of subsequent retrieval results and lead to reduced accuracy. Consequently, a data chunker is introduced to split the clean data into different chunks with clearer semantic meaning. For the \textbf{chunking strategy}, we chose to store each question on a private ruling webpage as a single data chunk, rather than storing the entire private ruling page information as a single data chunk. Each question is represented as embeddings stored in the vector database. For \textbf{metadata storage}, an approach in LangChain's implementation, where metadata is stored within the vector database, facilitates efficient retrieval alongside embeddings, reducing the need for additional relational database queries and minimising indexing overhead \cite{LangChain_2023}. This aligns with industry best practices, as seen in Pinecone's recommendation for storing metadata within vector databases when fast retrieval and contextual enrichment are key priorities \cite{Pinecone_BestPractices}. Therefore, we decided to store the question-related information within the vector database metadata. When the copilot retrieves relevant results, the metadata is fetched simultaneously with the question embeddings, ensuring efficient contextual retrieval.
    \item \textbf{Vector database}: When evaluating the choice between a \textbf{vector database (DB) vs. a knowledge graph (KG)}, although a KG may capture relationship between entities, such as questions and rules more effectively, it is more cost-efficient to store plain text in a vector database. Considering the ruling data is updated on a weekly basis, it is costly to correctly and efficiently capture the entities and relationships, and create the KGs. For instance, constructing KGs often requires significant manual effort and poses challenges for frequent updates \cite{Ontoforce_2024}. Additionally, integrating data from diverse sources into KGs can be a long, manual process, making seamless integration difficult \cite{Coveo_2023}. Therefore, we decided to proceed with the vector database for now, balancing cost and complexity.
    \item \textbf{Reranker}: Once the \textbf{retriever} generates an initial list of ranked results, the \textbf{reranker} can further evaluate their relevance by using the LLM-based \textbf{evaluator} to refine the ranking. This may increase latency and cost due to additional tokens processed by the LLM. However, integrating a reranker significantly enhances accuracy, as rerankers are much more accurate than embedding models \cite{Pinecone_Rerankers}. Additionally, rerankers can enhance the relevance of retrieved information in a RAG system, providing more accurate and contextually relevant results \cite{Zilliz_Rerankers}. Therefore, we decided to incorporate the \textbf{reranker} to the architecture design.
\end{itemize}

\subsubsection{Architecture Analysis for Adaptability Context Scenario}
We identified a gap in the architecture regarding adaptability, as no approaches were supporting it. We decided to include a \textbf{memory} component to store the user feedback for each query's response, enabling the copilot to learn and adapt over time. The user profile stored in \textbf{memory} can be refined base on feedback, enabling the copilot to tailor responses more effectively. Additionally, the copilot can dynamically select response templates based on the updated user profile, improving personalisation and accuracy in future interactions.

\subsubsection{Architecture Analysis for Efficiency Context Scenario}
We found a lack of architecture support for the efficiency context scenario. We decided to add logic in the \textbf{reasoning and planning} to first check whether similar queries have been previously answered in the memory. If a match is found, the copilot will directly return the previously response to the tax professional to improve efficiency and reduce cost associated with additional computation and LLM queries. To ensure accuracy, the copilot will also verify that the response aligns with the most up-to-date legal information, avoiding outdated recommendations when laws/regulations have changed.

\subsubsection{Architecture Analysis for Transparency Context Scenario}
The \textbf{generator} component in the previous architecture already includes reasoning behind the recommendation. To further improve transparency and assist a tax professional in understanding how the copilot arrived at its recommendation, while also reminding them that the recommendation is AI-generated, we decided to include a detailed list of all legal sources referenced by the copilot (e.g. IDs and links of tax rulings stored in the chunk metadata) and specify the version of the LLM in the response template.

\subsubsection{Architecture Analysis for Observability Context Scenario}
The previous architecture did not provide support for observability. To address this, we decided to introduce an \textbf{AgentOps infrastructure} layer that cross-cuts all copilot components. This layer will log all queries, responses, user feedback, and the input and output of each component, providing comprehensive visibility into the copilot's operations and improving \textbf{monitoring} and \textbf{evaluation} capabilities. Regarding \textbf{using ground truth or not}, we decided to gather a set of ground truth data from ATO open forum's questions and ATO certified replies. For complex scenarios, we chose to initially collect ground truth data directly from domain experts, then use LLM to generate more ground truth data for evaluation.

\subsubsection{Architecture Analysis for Contestability Context Scenario}
In the previous design, tax professionals were able to provide scores and feedback on the recommendation generated by the copilot. The improved architecture now includes a \textbf{memory} component that not only stores user feedback but also tracks the resolution process.

\subsubsection{Architecture Analysis for Privacy Context Scenario}
\textbf{Guardrail scope} is an important design decision. The previous architecture included a desensitiser to ensure that private information is not disclosed in the user prompt. However, sensitive information could also be present in the input and output of other components. Thus, the new architecture introduces \textbf{guardrails} that are applied across all components. The guardrails can also be extended to cover additional concerns such as relevance and fairness.

\section{Conclusion}
In this paper, we present AgentArcEval, a novel architecture evaluation method specially designed to address the unique characteristics of agents. Additionally, we introduced a catalogue of agent-specific general scenarios to guide the generation of context scenarios for specific agents. Through a real-world case study on the architecture evaluation of the Luna tax copilot, we demonstrated the usefulness of AgentArcEval in addressing the architectural complexities of agents. We are currently applying AgentArcEval to additional FM-based agentic systems and will report comparative results in future work. 

Our near-term goal is to  establish an evolvable community-driven approach to the evaluation of agent systems by publishing the method and catalogue as a living document and inviting contributions from the broader community.




\end{document}